\documentclass[journal=jpcbfk,manuscript=article]{achemso}
\usepackage{longtable}
\usepackage{multirow}
\usepackage{amsmath}
\usepackage{subfigure}
\usepackage[usenames,dvipsnames]{color}
\usepackage{soul}
\usepackage{threeparttable}
\usepackage{xr}
\usepackage{graphicx}
\usepackage{amsmath,amssymb}
\usepackage{caption}
\usepackage{color}
\usepackage{dcolumn}
\usepackage{bm}
\usepackage{float}
\usepackage{subfig}
\usepackage{textcomp}
\usepackage{url}
\usepackage[normalem]{ulem}

\makeatletter

\author{Seyedeh Maryam Salehi} \affiliation[University of
  Basel]{Department of Chemistry, University of Basel,
  Klingelbergstrasse 80 , CH-4056 Basel, Switzerland.}
\author{Markus Meuwly} \affiliation[University of Basel]{Department of
  Chemistry, University of Basel, Klingelbergstrasse 80 , CH-4056
  Basel, Switzerland.}  \email{m.meuwly@unibas.ch}

\title {Cross-Correlated Motions in Azidolysozyme}

\begin{document}

\date{\today}

\begin{abstract}
The changes in the local and global dynamics of azide-labelled
Lysozyme compared with that of the wild type protein are
quantitatively assessed for all alanine residues along the polypeptide
chain. Although attaching -N$_3$ to alanine residues has been
considered to be a minimally invasive change in the protein it is
found that depending on the location of the Alanine residue the local
and global changes in the dynamics differ. For Ala92 the change in the
cross correlated motions are minimal whereas attaching -N$_3$ to Ala90
leads to pronounced differences in the local and global correlations
as quantified by cross correlation coefficients of the C$_{\alpha}$
atoms. It is also demonstrated that the spectral region of the
asymmetric azide stretch distinguishes between alanine attachment
sites whereas changes in the low frequency, far-infrared region are
less characteristic.
\end{abstract}

\section{Introduction}
To characterize cellular processes at a molecular level the structure
and dynamics of proteins needs to be understood. Such knowledge is
also valuable to direct development and improvement of
pharmaceutically active ligands in drug design
efforts.\cite{schuler:2017,zhou:2016,zhang:2019} Optical spectroscopy
is one possibility to capture the structural and functional dynamics
of proteins in the condensed phase. One of the great challenges is to
probe site-selective dynamics. This is required in order to
specifically target protein sites that are responsible for function.\\

\noindent
During the past 10 years a range of non-natural small molecule
modifications to proteins has been proposed. They include - but are
not limited to - attaching nitrile to amino acids\cite{gai:2003},
using the sulfhydryl band of cysteines,\cite{hamm:2008},
cyano\cite{romesberg.cn:2011} groups, nitrile
labels,\cite{fayer.ribo:2012} complexation with
SCN\cite{bredenbeck:2014}, cyanophenylalanine\cite{thielges:2015}, or
cyanamide.\cite{cho:2018} For Lysozme, ruthenium carbonyl complexes
have also been shown to provide an understanding of the water dynamics
from 2d-infrared experiments.\cite{kevin:2012,kevin.2:2012,kevin:2014}
In this case it has been explicitly demonstrated that dynamic
hydration extends over distances 20 \AA\/ away from the protein
surface, consistent with recent simulations on hydrated
Hb.\cite{MM.hb:2018,MM.hb:2020}\\

\noindent
Based on recent studies\cite{MS.lys:2021} the vibrational dynamics of
N$_3^{-}$ in the gas phase and in solution can be captured
quantitatively.\cite{MS.n3:2019} Moreover, azidoalanine (AlaN$_3$) is
one of ideal labels that has been shown to be positionally sensitive
probe for the local dynamics.\cite{MS.lys:2021} AlaN$_3$ has a
comparatively large extinction coefficient and it absorbs around $\sim
2100$ cm$^{-1}$. The incorporation of N$_3^{-}$ to alanine (Ala) is
technically feasible and leads to small
perturbations.\cite{bertozzi:2002} Thus, AlaN$_3$ as also shown
previously\cite{MS.lys:2021}, is an ideal modification to study the
local protein dynamics.\\

\noindent
In the previous work, N$_3^-$ was attached to all Ala residues in
Lysozyme with the aim to determine changes in the local dynamics
around the modification sites and effects on the global dynamics of
the protein. First, the methods are discussed. Next, results on the
root mean squared fluctuations and the dynamical cross correlation
maps are presented. finally, the spectroscopy in the low-frequency and
around the asymmetric azide stretch vibration is considered and
conclusions are drawn.\\

\section{Methods}
\subsection{Molecular Dynamics Simulation}
\begin{figure}[H]
\begin{center}
\includegraphics[width=0.8\textwidth]{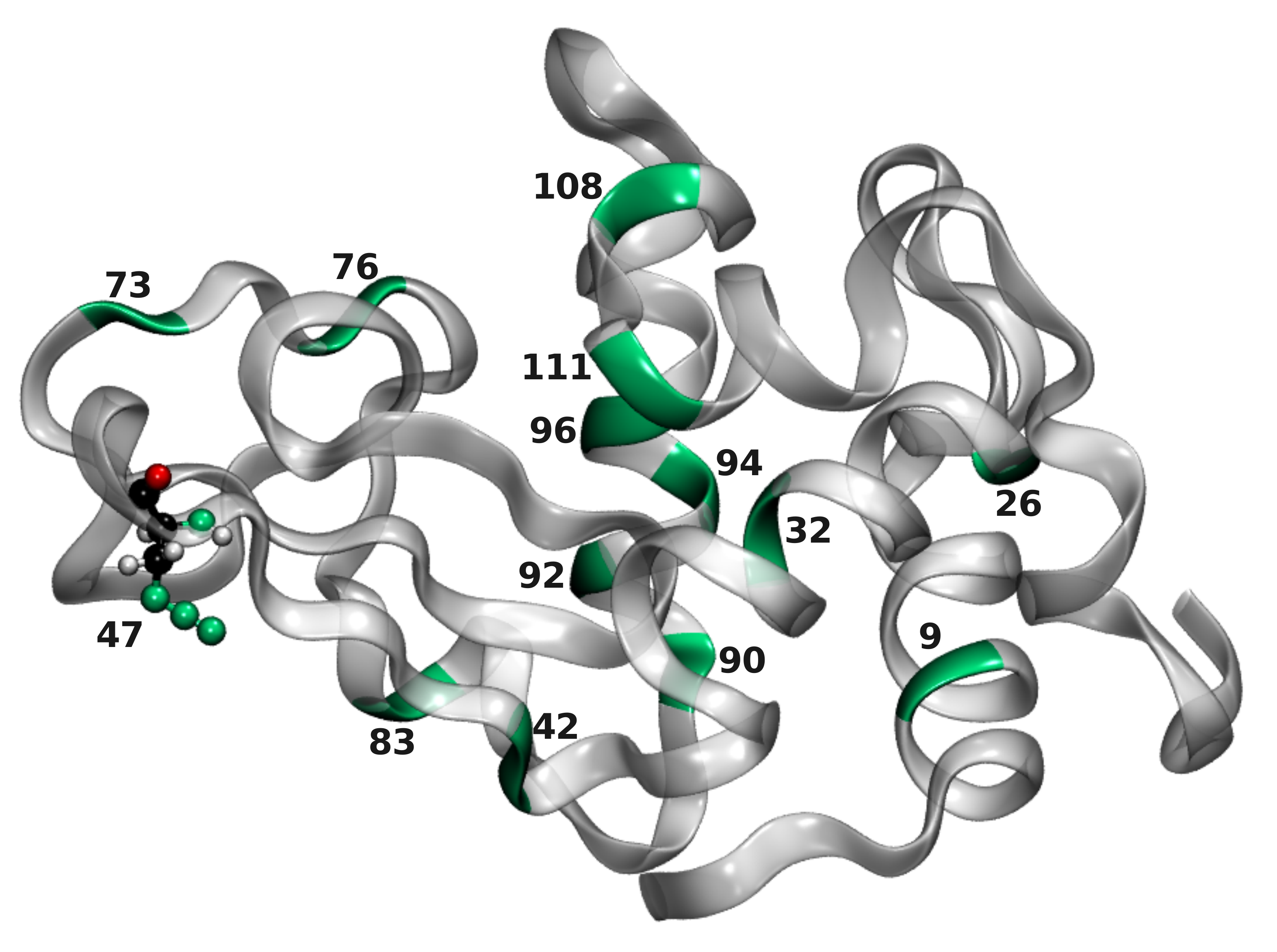}
\caption{Lysozyme structure with indicated alanine residues at
  positions 9, 26, 32, 42, 47, 73, 76, 83, 90, 92, 94, 96, 108,
  111. Ala residues are displayed as green NewRibbons while the rest
  of the protein structure is shown with white NewRibbons. Ala47N$_3$
  is demonstrated in CPK as an example of alanine modified residue.}
\label{fig:lys}
\end{center}
\end{figure}

Molecular Dynamics (MD) simulations of WT and all the modified
AlaN$_3$ labels is done using CHARMM\cite{charmmFF22} force field. For
the simulations with multi-dimensional RKHS PES, an interface is
written.\cite{MS.n3:2019} MD simulations are performed with TIP3P
water \cite{TIP3P-Jorgensen-1983} model in a cubic box of size
$(62.1)^3$ \AA\/$^3$. Figure \ref{fig:lys} represents the Lysozyme
structure used in the current study with all modified Ala
residues. The initial x-ray structure corresponds to WT human Lysozyme
(3FE0\cite{pdblyso-2009}).  First, the system is minimized followed by
heating and equilibration for 100 ps. Then, the production runs of 2
ns are carried out for all 14 AlaN$_3$ labels using $NVT$ ensemble and
snapshots for analysis were recorded every 5 fs.  Bond lengths
involving H-atoms were constrained using the
SHAKE\cite{SHAKE-Gunsteren-1997} algorithm and all non-bonded
interactions were evaluated using shifted interactions with a cutoff
of 14 \AA\/ switched at 10 \AA\/.\cite{Steinbach1994}\\

\subsection{Dynamical Cross Correlation Maps}
To quantitatively determine the effect of ligand binding on the
protein dynamics, dynamical cross correlation
maps\cite{karplus:1991,ornstein1997} (DCCM) and difference dynamical
cross correlation maps ($\Delta$DCCM) are calculated using the Bio3D
package.\cite{bio3d}. Dynamic cross-correlation matrices with
coefficients
\begin{equation}
 C_{ij} = \langle \Delta r_{i}. \Delta r_{j}\rangle / (\langle \Delta
 r_{i}^{2} \langle \Delta r_{j}^{2}\rangle)^{1/2}
\label{eq:dccm}
\end{equation}
were determined from the positions of the main chain C$_{\alpha}$
atoms in amino acids $i$ and $j$ with positions $r_{i}$ and $r_{j}$.
$\Delta r_{i}$ and $\Delta r_{j}$ determine the displacement of the
$i$th C$_{\alpha}$ from its mean position over the entire trajectory.
For DCCM calculation, the Bio3D package\cite{bio3d} is used. Note that
DCCMs describes the correlated and anti-correlated motions in a
protein whereas differences $\Delta$DCCM report on pronounced
differences between unmodified and modified proteins.

\subsection{Infrared Spectroscopy}
The infrared spectrum is calculated by the Fourier transform over the
dipole moment autocorrelation function. To that end, the dipole moment
$\vec{\mu}$ is obtained for the entire protein and -N$_3$ label
separately from the simulation trajectories of 2 ns production
run. For the IR spectra the correlation function $C(t) = \langle
\vec{\mu}(0) \vec{\mu}(t) \rangle$ was determined from snaphots saved
every 5 fs. Then, the fast Fourier transform of $C(t)$ was determined
using a Blackman filter and the result was multiplied using a quantum
correction factor $\beta \hbar \omega / (1 - \exp{(-\beta \hbar
  \omega)})$ where $\beta = 1/(k_{\rm B} T)$.\cite{tavan:2004}

\section{Results}
\section{Global Dynamics (RMSF and DCCM)}
\begin{figure}[H]
\begin{center}
\includegraphics[width=\textwidth]{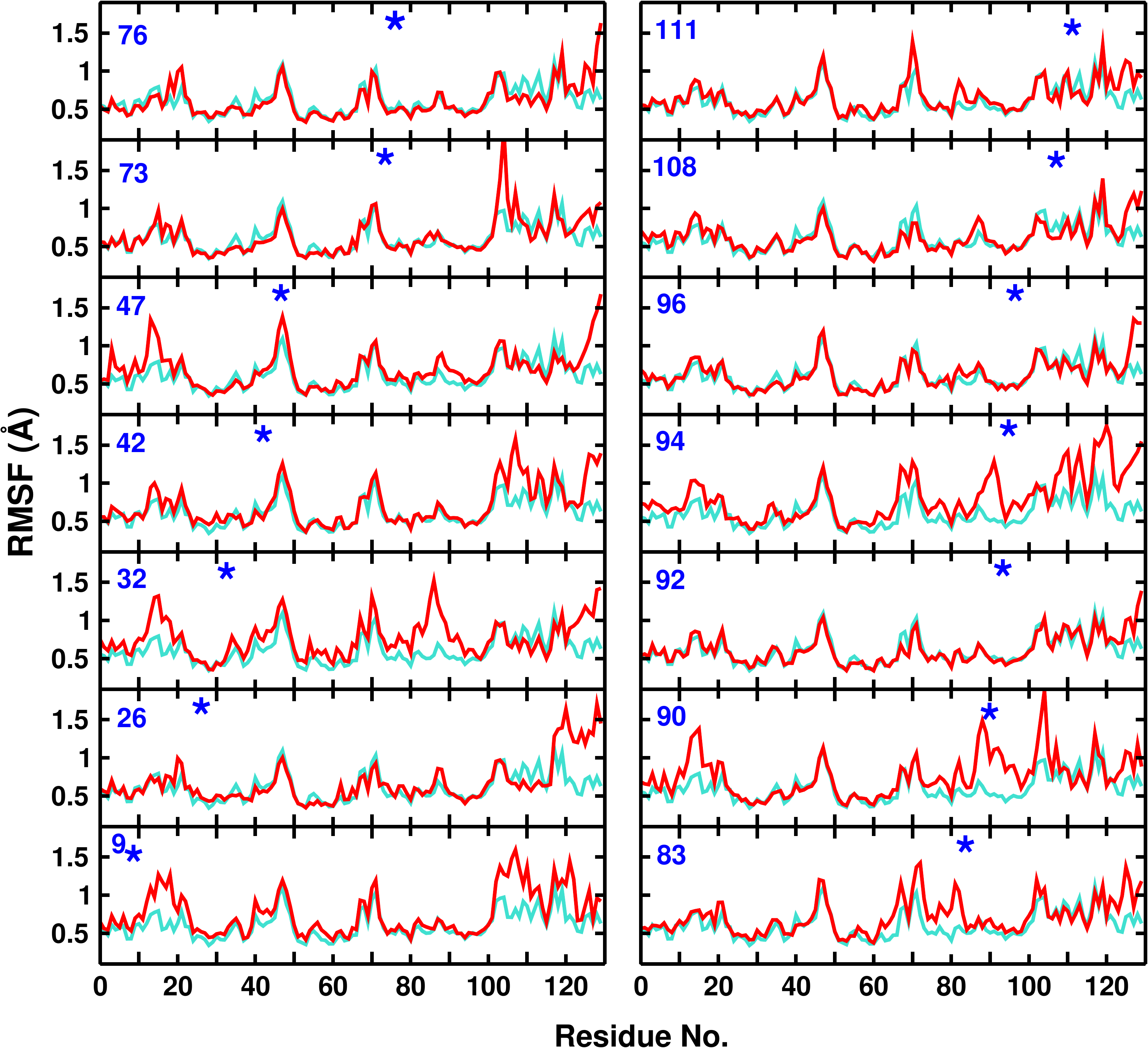}
\caption{Root mean squared fluctuations for the C$_{\alpha}$ atoms of
  the WT (turquoise) protein and including different AlaN$_3$ (red)
  modification sites. The position of the modified residue is
  indicated by an asterisk.}
\label{fig:rmsf}
\end{center}
\end{figure}

To probe the flexibility or rigidity of the unmodified and modified
proteins the root mean squared fluctuation (RMSF) of lysozyme in
solution before and after modification is compared with original X-ray
structure as the reference. The comparison is based on all
C$_{\alpha}$ atoms in the protein and the result is shown in Figure
\ref{fig:rmsf} for all Ala residues. The RMSF changes of the modified
proteins compared with unmodified Lysozyme range from minor (Ala26,
Ala32, Ala42, Ala73, Ala76, Ala92, Ala96, Ala108, Ala111) to major
(Ala9, Ala47, Ala83, Ala90, Ala94), see Figure
\ref{fig:rmsf}. Moreover, attaching an -N$_3$ label to one site may
also lead to larger fluctuations at other neighbor or non-neighbor
residues. As an example, modification of Ala47 to Ala47N$_3$ leads to
larger RMSFs for residues 12 to 18, 45 to 49, 87 to 93, and 122 to
130.\\

\noindent
Another example for a more pronounced change in the local flexibility
concerns modification of Ala32. In this case the region for residues
80 to 100 becomes considerably more flexible which can also be
explained by the close contact of the two helices Ala32 on the one
hand and residues 90 to 100 are part of, see Figure
\ref{fig:lys}. Finally, attaching N$_3$ to Ala94 leads to a range of
changes in the flexibility of nearby and more distant residues,
including residues [10-20], [60-70], [85-100], and [110-130]. It is
noteworthy to remark that in general, attaching -N$_3$ leads to
increased flexibility of the protein and that often the terminus
around residue 130 becomes more flexible. One example for which the
flexibility is decreased concerns modification at Ala108 for which the
region [65-75] becomes more rigid.\\

\begin{figure}[H]
\begin{center}
\includegraphics[width=0.9\textwidth]{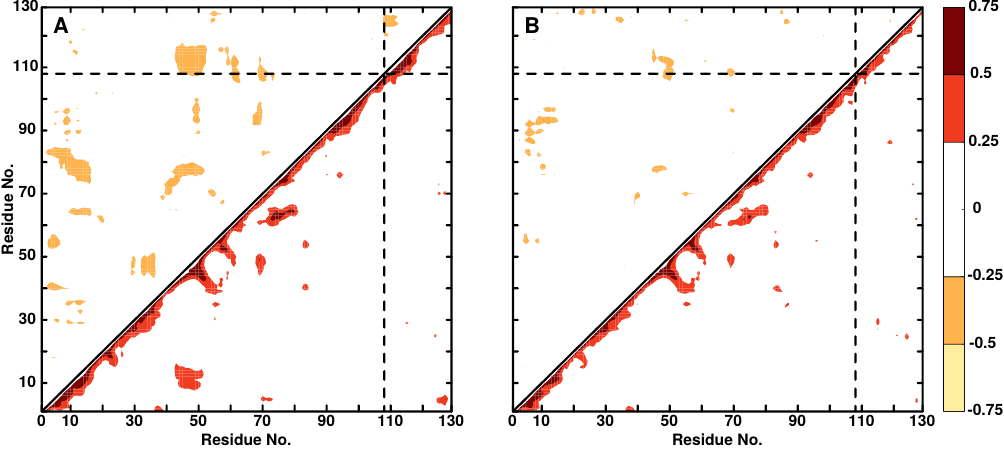}
\caption{DCCM for WT lysozyme (panel A) and Ala108N$_3$ (panel
  B). Positive correlations are in the lower right triangle, negative
  correlations in the upper left triangle. Only correlation
  coefficients with an absolute value greater than 0.25 are
  displayed.}
\label{fig:dccmwt-108-diff}
\end{center}
\end{figure}

\noindent
\textbf{Dynamical Cross Correlation Maps:} DCCMs describe the
correlated ($C_{ij} \sim 1$) and anti-correlated ($C_{ij} \sim -1$)
motions within a protein. As an example, the DCCMs for the WT lysozyme
and Ala108N$_3$ are shown in Figure \ref{fig:dccmwt-108-diff}. Bulges
along the diagonal correspond to helices whereas features extending
away from the diagonal orthogonal to it are $\beta-$sheet
structures. For the WT there is a hinge motion at the intersection
between residues 10 and 50 with $0.25 \leq C_{ij} \leq 0.5$ which
disappears upon modification at position Ala108. The intensity of such
hinge motions should be compared with insulin monomer for which
disulfide bonds are characterized by $0.55 \leq C_{ij} \leq 0.70$,
i.e. about a factor of two more intense than in the present
case. Similarly, there are several anticorrelated motions in the WT
protein, e.g. involving residues 10 and 80 or 50 and 110 which largely
disappear upon modification at Ala108. It is also instructive to
compare the DCCM with the RMSF in Figure \ref{fig:rmsf}. As was
already discussed, the RMSF for Ala108N$_3$ is reduced for residues in
the range [65,75] compared with the WT protein. This can also be seen
in the DCCM for which the positive correlation involving residues
[70-80,60] is considerably stronger in the WT compared with the
modified protein.\\

\subsection{Local dynamics ($\Delta$DCCM)}
To further analyze the modification sites, difference maps
$\Delta$DCCM are calculated for all the AlaN$_3$ modifications which
report on residues with pronounced changes in the dynamics after
modification. The range of differences considered includes changes
($>\lvert 0.25 \rvert$) and smaller than ($>\lvert 0.75 \rvert$). For
differences $|\Delta C_{ij}| < 0.25$ the changes are too insignificant
whereas values $|\Delta C_{ij}| > 0.75$ were not observed.\\

\begin{figure}[H]
\begin{center}
\includegraphics[width=0.85\textwidth]{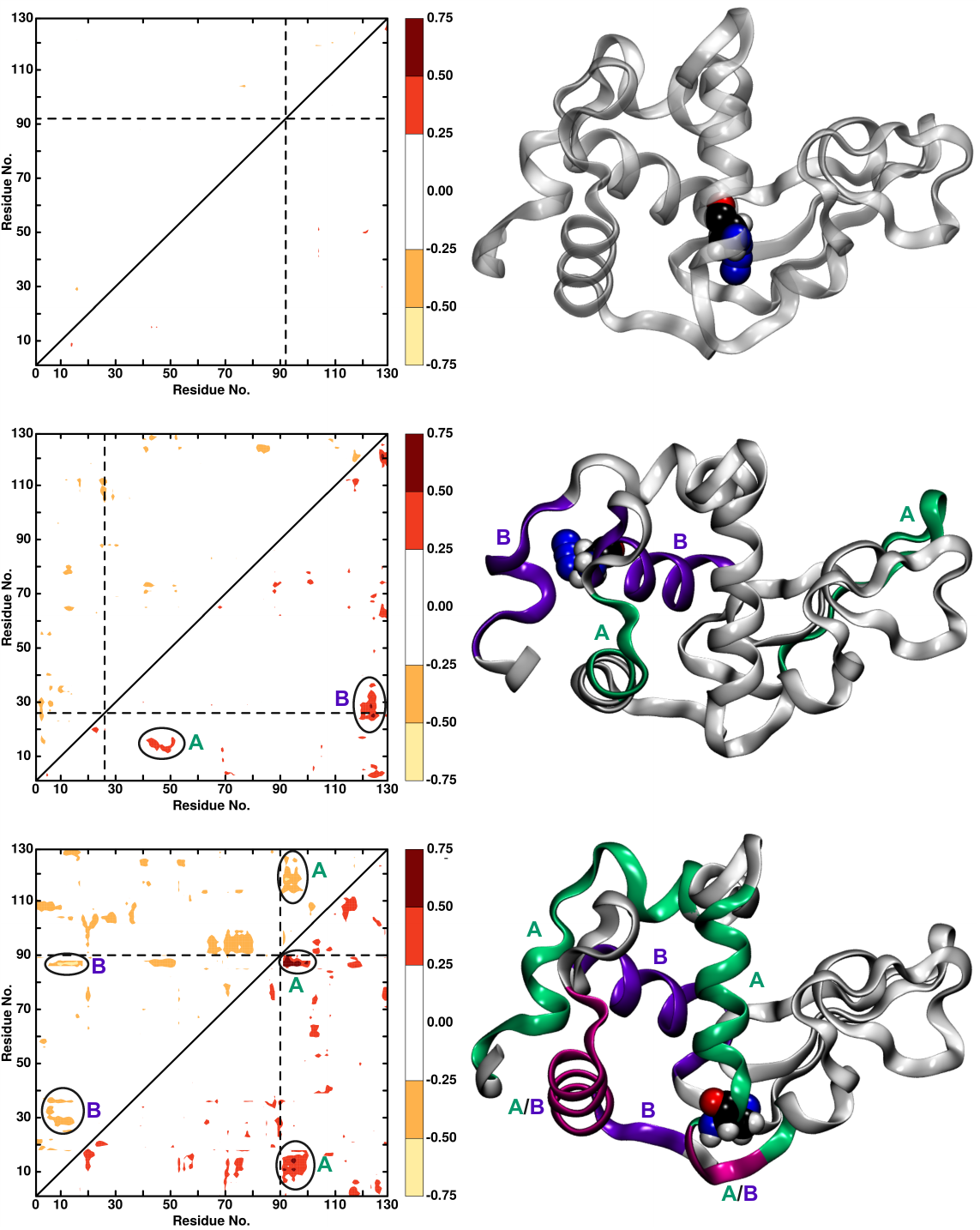}
\caption{Difference dynamic cross correlation maps ($\Delta$DCCM)
  between WT and Ala92N$_3$, Ala26N$_3$, Ala90N$_3$ (top to
  bottom). Positive differences are in the lower right triangle,
  negative differences in the upper left triangle. Only differences
  with an absolute value greater than 0.25 are displayed. Right panels
  show the protein structure with features A (green) and B (violet)
  and common residues between feature A and B (magenta) of
  $\Delta$DCCM plots highlighted in color. The AlaN$_3$ label is shown
  in VDW.}
\label{fig:diff92-26-90}
\end{center}
\end{figure}

\noindent
Figure \ref{fig:diff92-26-90} shows the corresponding $\Delta$DCCM
maps for residues Ala92N$_3$, Ala26N$_3$, and Ala90N$_3$ as examples
for minor, medium and major effects on the protein dynamics after
modification. Incorporation of -N$_3$ into the protein at position
Ala92 leads to insignificant changes in the correlation between the
residues (Figure \ref{fig:diff92-26-90} top panel). This is consistent
with the findings from analysis of the RMSF, see Figure \ref{fig:rmsf}
, which shows only minor variation in the fluctuations without and
with N$_3$ attached to Ala92. On the contrary, for Ala26N$_3$
ligand-induced effects between residues $[12,18]$ and $[42,52]$
(feature A) and residues $[23,35]$ and $[120,127]$ (feature B) are
found, see Figure \ref{fig:diff92-26-90} middle panel. Again, these
changes can also be found in the RMSF analysis in Figure
\ref{fig:rmsf} although there, the effects specifically for residues
    [12,18] and [23,35] are smaller, albeit still visible. Both, local
    (feature B) and global (feature A) changes in the protein dynamics
    are found.\\

\noindent
Finally, the difference map for Ala90N$_3$ (Figure
\ref{fig:diff92-26-90} bottom panel) shows strong effects on the
dynamics and couplings between residues after attaching N$_3$ to the
Alanine residue. Feature A indicates coupling between residues
$[90,102]$ and $[5,18]$/$[85,90]$/$[113,128]$ whereas feature B refers
to coupled residues $[3,18]$ and $[87,89]/[27-38]$. It is noted that
residues $[5,18]$ and $[87,89]$ are common between feature A and
B. These findings suggest that residues couple both locally and
through space. Again, residues involved in features A and B belong to
those with the largest variations in the RMSF, see Figure
\ref{fig:rmsf}. For other $\Delta$DCCM representations, see Figure
\ref{sifig:diff9} to \ref{sifig:diff111}.\\

\subsection{Spectroscopy}
It is also of interest to further consider the vibrational
spectroscopy for the different modified proteins. In particular, it is
of interest whether attaching the azide label at different positions
along the polypeptide chain leads to discernible changes only in the
asymmetric N$_3$ stretch vibration or whether the spectroscopy in the
low-frequency (far infrared, THz) range is also expected to be affected
by the modification.\\

\begin{figure}[H]
\begin{center}
\includegraphics[width=0.5\textwidth,angle=-90]{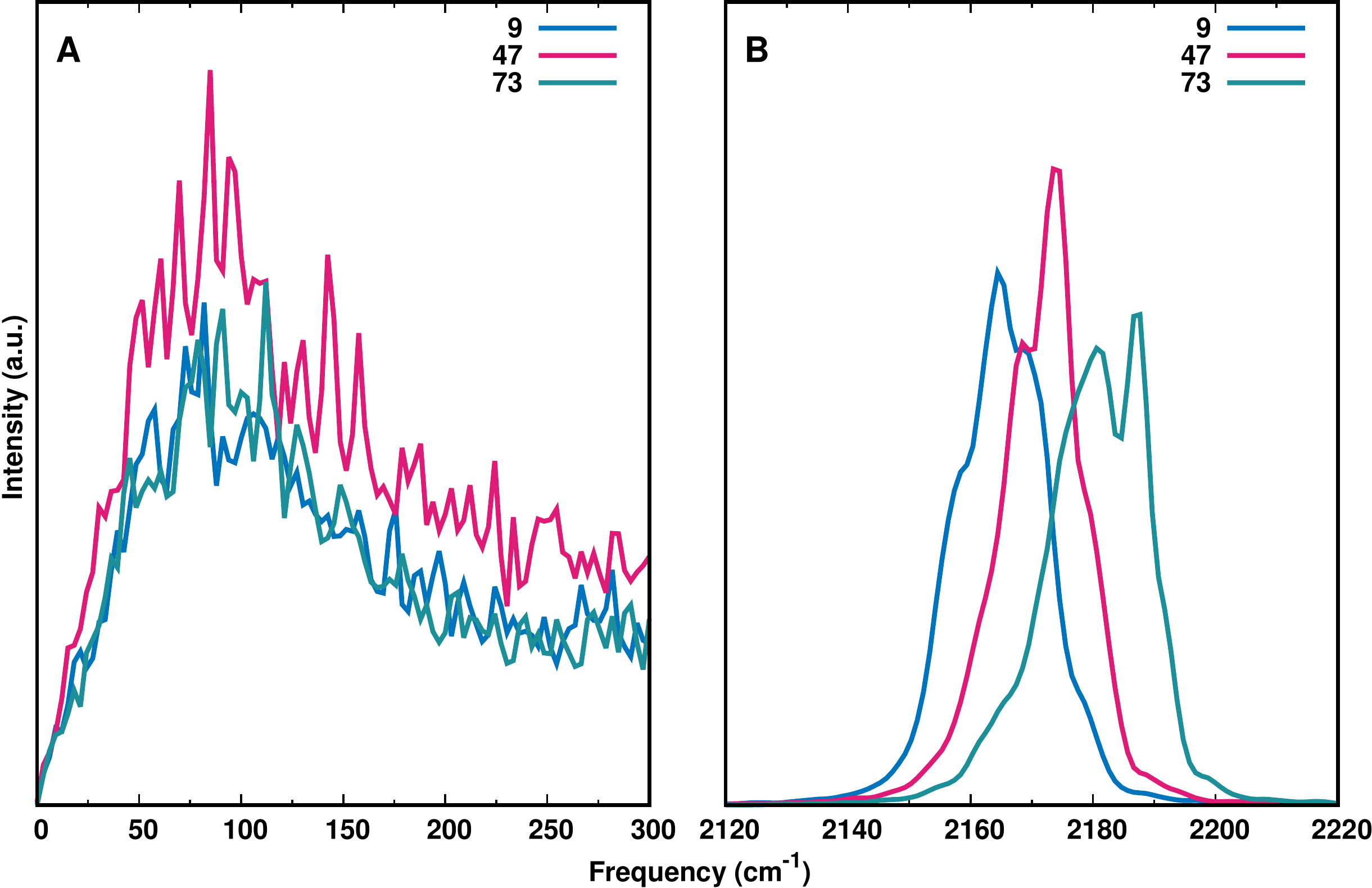}
\caption{IR spectrum from Fourier Transform of the dipole moment
  auto-correlation function of the entire protein (panel A) and the
  -N$_3$ label (panel B) for AlaN$_3$ modifications at positions 9
  (blue), 47 (red), and 73 (green). The results show that the region
  around the asymmetric N$_3$ stretching vibration is considerably
  more discriminating than the THz region of the spectrum.}
\label{fig:ir}
\end{center}
\end{figure}

\noindent
Figure \ref{fig:ir}A reports the IR spectrum in the low frequency
range (0-300 cm$^{-1}$) with modifications at positions 9, 47, and 73
together with the high frequency range (2130-2230 cm$^{-1}$, panel B)
which captures the asymmetric stretch of the -N$_3$ label. The IR
spectrum for all AlaN$_3$ modifications is shown in Figure
\ref{sifig:ir}. Using the same energy function for all AlaN$_3$
moieties, the results demonstrate that the IR spectra differ in terms
of the position of the frequency maxima and their full widths at half
maximum (FWHM). Compared with earlier work which determined the 1-d IR
lineshape from the frequency fluctuation correlation function
(FFCF)\cite{MS.lys:2021} the present analysis confirms that the
frequency maxima of the most red (Ala9) and most blue (Ala73) shifted
azide vibrations differ by $\sim 15$ cm$^{-1}$.\\

\noindent
In addition, the spectrum in the low-frequency region (0 to 300
cm$^{-1}$) is reported in Figure \ref{fig:ir}A. The FFCFs determined
previously reported pronounced oscillations for -N$_3$-modification at
Ala9, Ala32, Ala42, and Ala92 but their origin remains
unclear. Similarly, earlier work in the region of the antisymmetric
stretch of the N$_3^{-}$ anion in ternary complexes with formic acid
dehydrogenase and NAD$^+$ and NADH, respectively, also reported
oscillations on the picosecond time scale of the
FFCF.\cite{cheatum:2016} Importantly, in this earlier work the azide
anion is not covalently bound to either the protein or the ligand but
rather replaces the formate reactant as a transition state analogue in
the active site of the protein. The reported oscillations occur on the
sub-picosecond to picosecond time scale. For the covalently bound
azide label the recurrences are rather on the sub-ps time scale
throughout. It was hypothesized that these recurrences potentially
originate from coupling of the azide label to low frequency modes of
the environment. However, the spectra in Figure \ref{fig:ir}A do not
support such an interpretation as irrespective of the position of the
azide label the spectral signatures in the range between 0 and 300
cm$^{-1}$ are largely identical.\\

\section{Conclusion}
The results discussed here confirm that azide attached to protein
alanine residues provides site-specific information about the
spectroscopy and dynamics. This is consistent with earlier findings
for Lysozyme and PDZ2..\cite{MS.lys:2021,stock:2018} Changes in
difference dynamical cross correlation maps range from insignificant
to major in that attaching -N$_3$ to Ala90 leads to major changes in
the overall protein dynamics through altered couplings across the
entire protein. On the other hand, this pronounced change in the
protein dynamics is not necessarily reflected in the infrared
spectroscopy of the azide label. Finally, it is demonstrated that the
spectral region of the asymmetric azide stretch discriminates much
more than the far infrared region which does not appear to exhibit
specific features depending on the location of the modification
site. The present work provides a molecularly resolved view of the
internal protein dynamics upon introducing small spectroscopic probes
at strategic positions of a protein.\\

\section*{Acknowledgments}
The authors gratefully acknowledge financial support from the Swiss
National Science Foundation through grant 200021-117810 and to the
NCCR-MUST.\\

\section*{Data Availability Statement}
The data that support the findings of this study are available from
the corresponding author upon reasonable request.

\bibliography{lyso}

\clearpage

\renewcommand{\thetable}{S\arabic{table}}
\renewcommand{\thefigure}{S\arabic{figure}}
\renewcommand{\thesection}{S\arabic{section}}
\renewcommand{\d}{\text{d}}
\setcounter{figure}{0}  
\setcounter{section}{0}  

\noindent
{\bf Supporting Information: Cross-Correlated Motions in
  Azidolysozyme}

\begin{figure}[H]
\begin{center}
\includegraphics[width=0.5\textwidth]{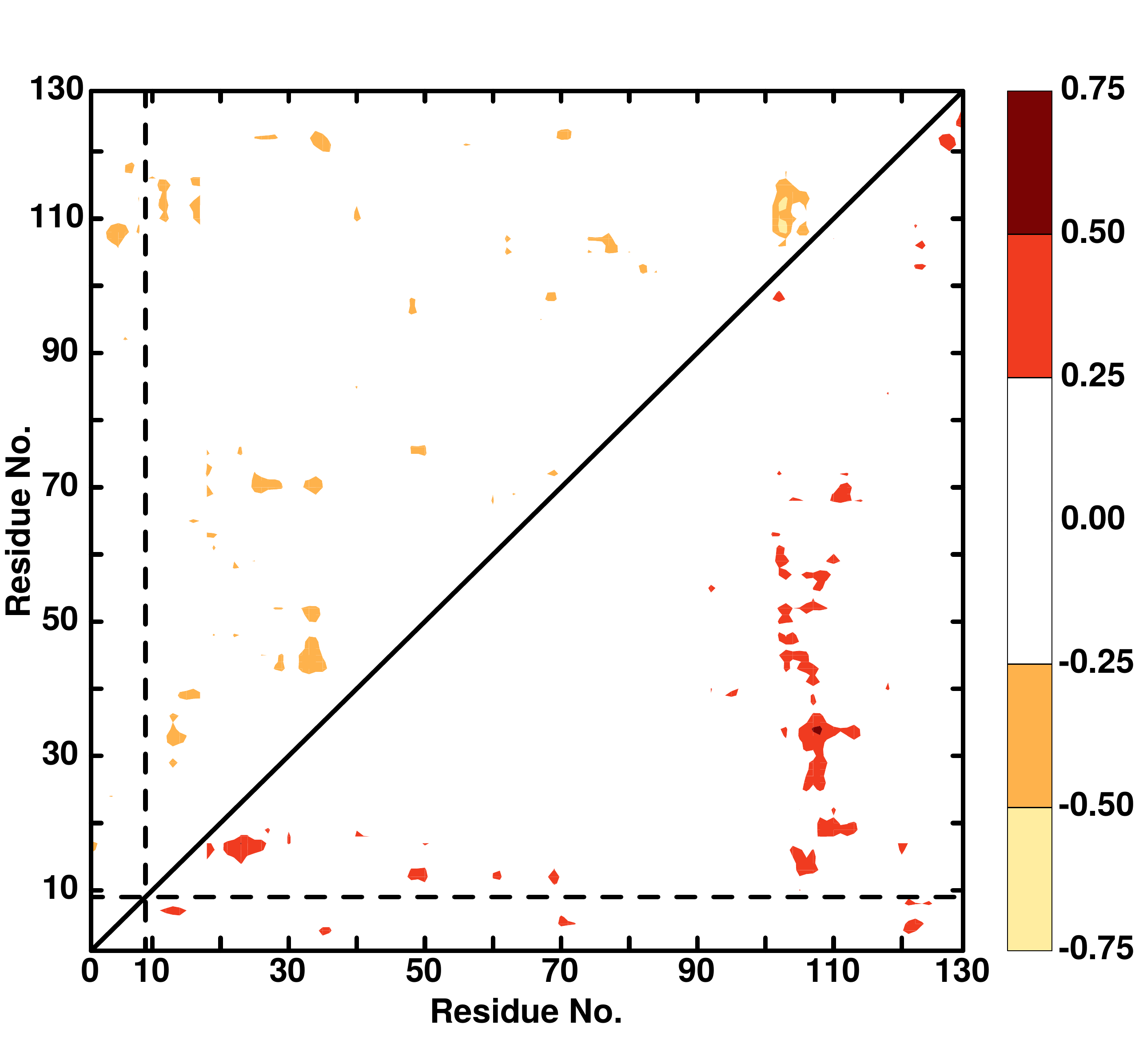}
\caption{$\Delta$DCCM between WT and Ala9N$_3$. Positive differences
  are in the lower right triangle, negative differences in the upper
  left triangle. Only differences with an absolute value greater than
  0.25 are displayed.}
\label{sifig:diff9}
\end{center}
\end{figure}

\begin{figure}[H]
\begin{center}
\includegraphics[width=0.5\textwidth]{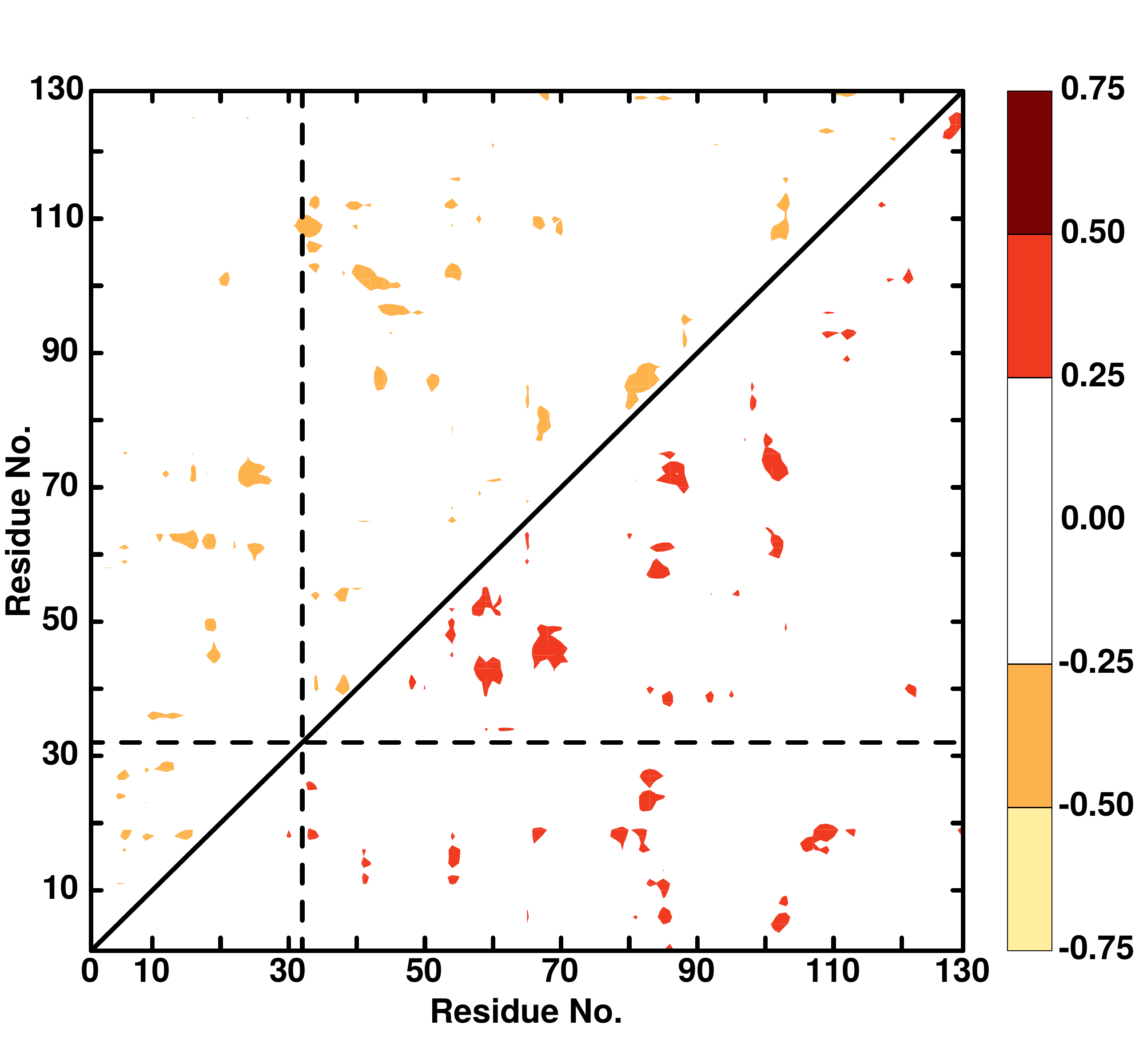}
\caption{$\Delta$DCCM between WT and Ala32N$_3$. Positive differences
  are in the lower right triangle, negative differences in the upper
  left triangle. Only differences with an absolute value greater than
  0.25 are displayed.}
\label{sifig:diff32}
\end{center}
\end{figure}

\begin{figure}[H]
\begin{center}
\includegraphics[width=0.5\textwidth]{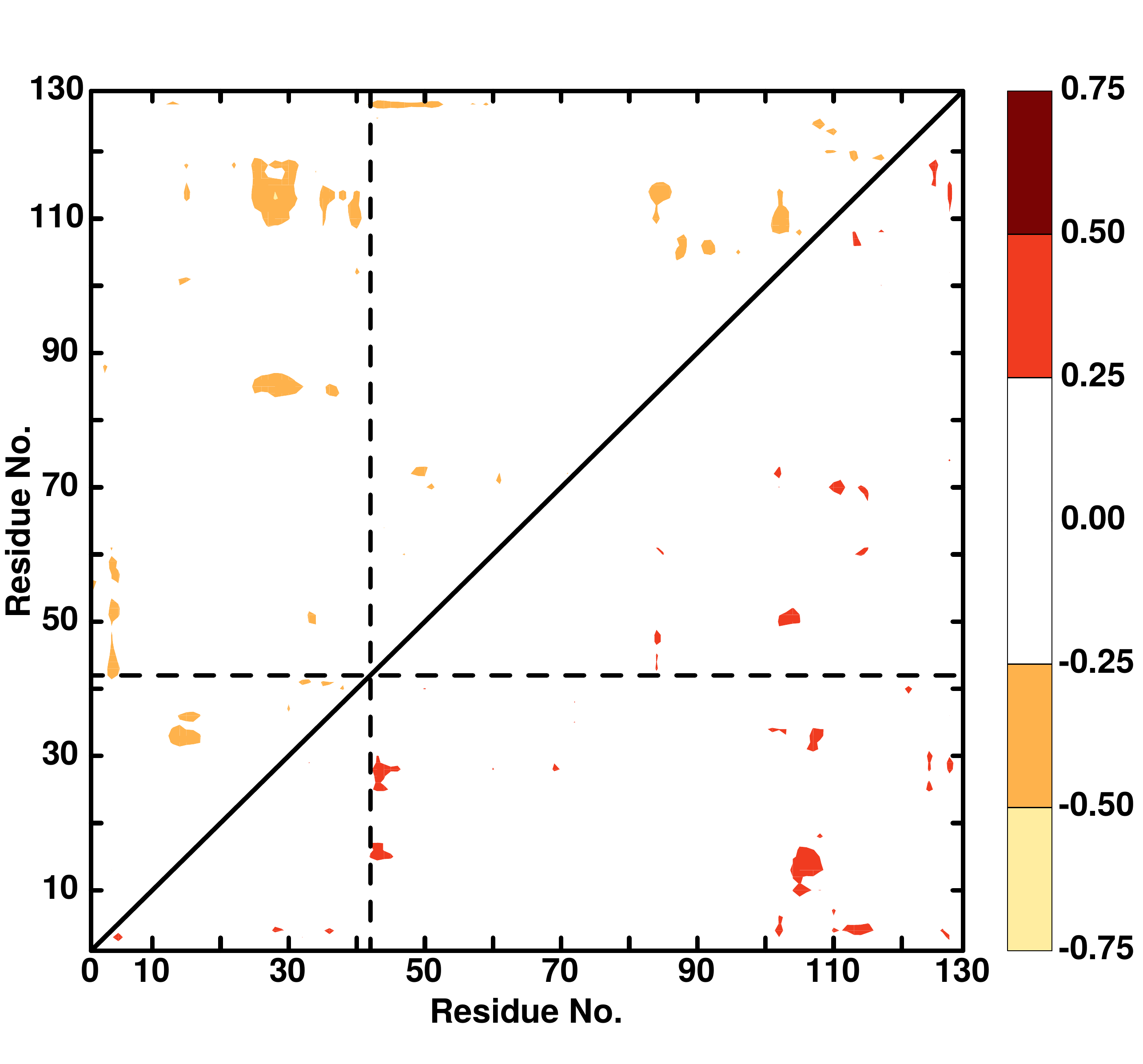}
\caption{$\Delta$DCCM between WT and Ala42N$_3$. Positive differences are in the lower right triangle, negative differences in the upper left triangle. Only differences with an absolute value greater than 0.25 are displayed.}
\label{sifig:diff42}
\end{center}
\end{figure}

\begin{figure}[H]
\begin{center}
\includegraphics[width=0.5\textwidth]{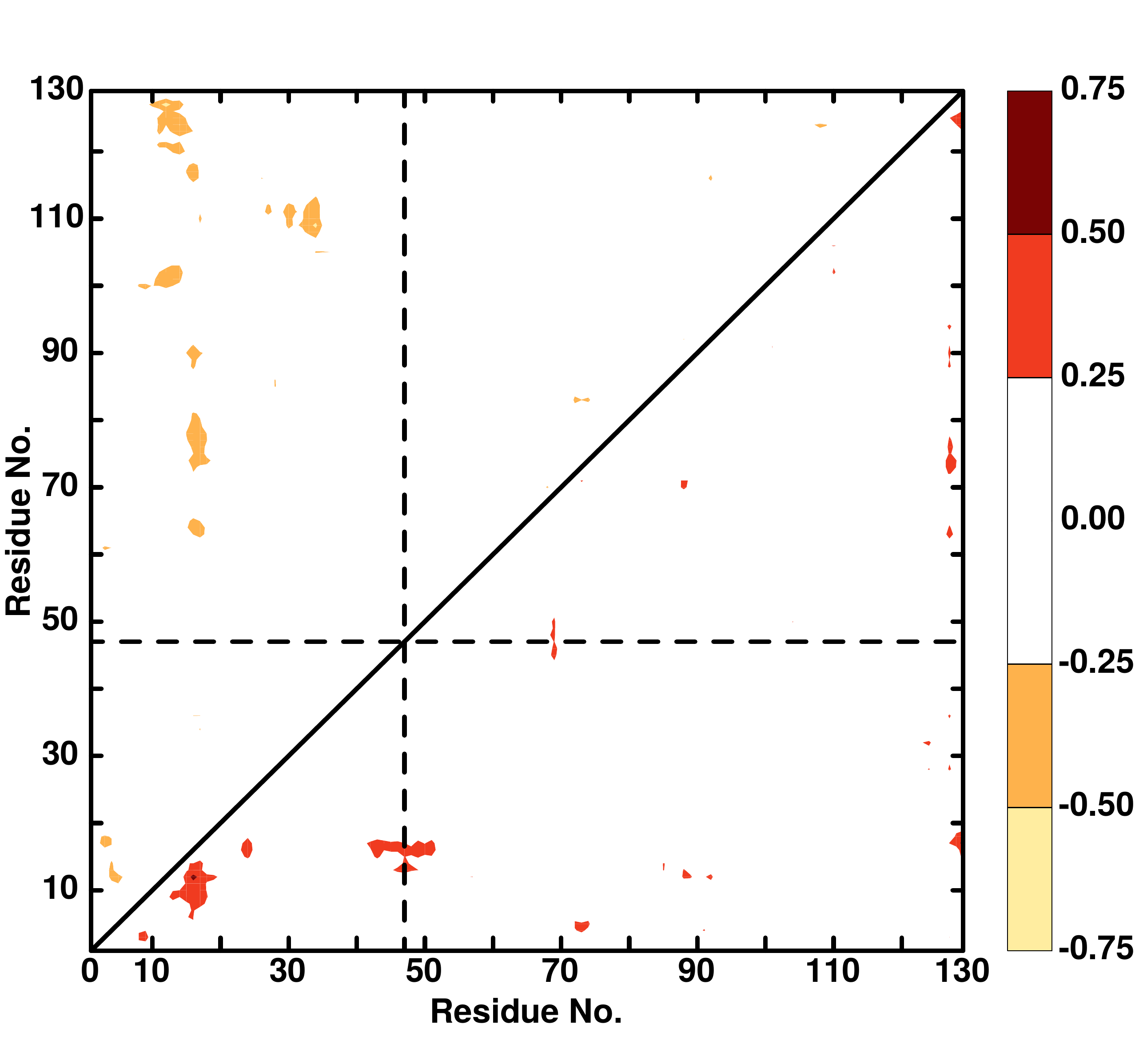}
\caption{$\Delta$DCCM between WT and Ala47N$_3$. Positive differences are in the lower right triangle, negative differences in the upper left triangle. Only differences with an absolute value greater than 0.25 are displayed.}
\label{sifig:diff47}
\end{center}
\end{figure}

\begin{figure}[H]
\begin{center}
\includegraphics[width=0.5\textwidth]{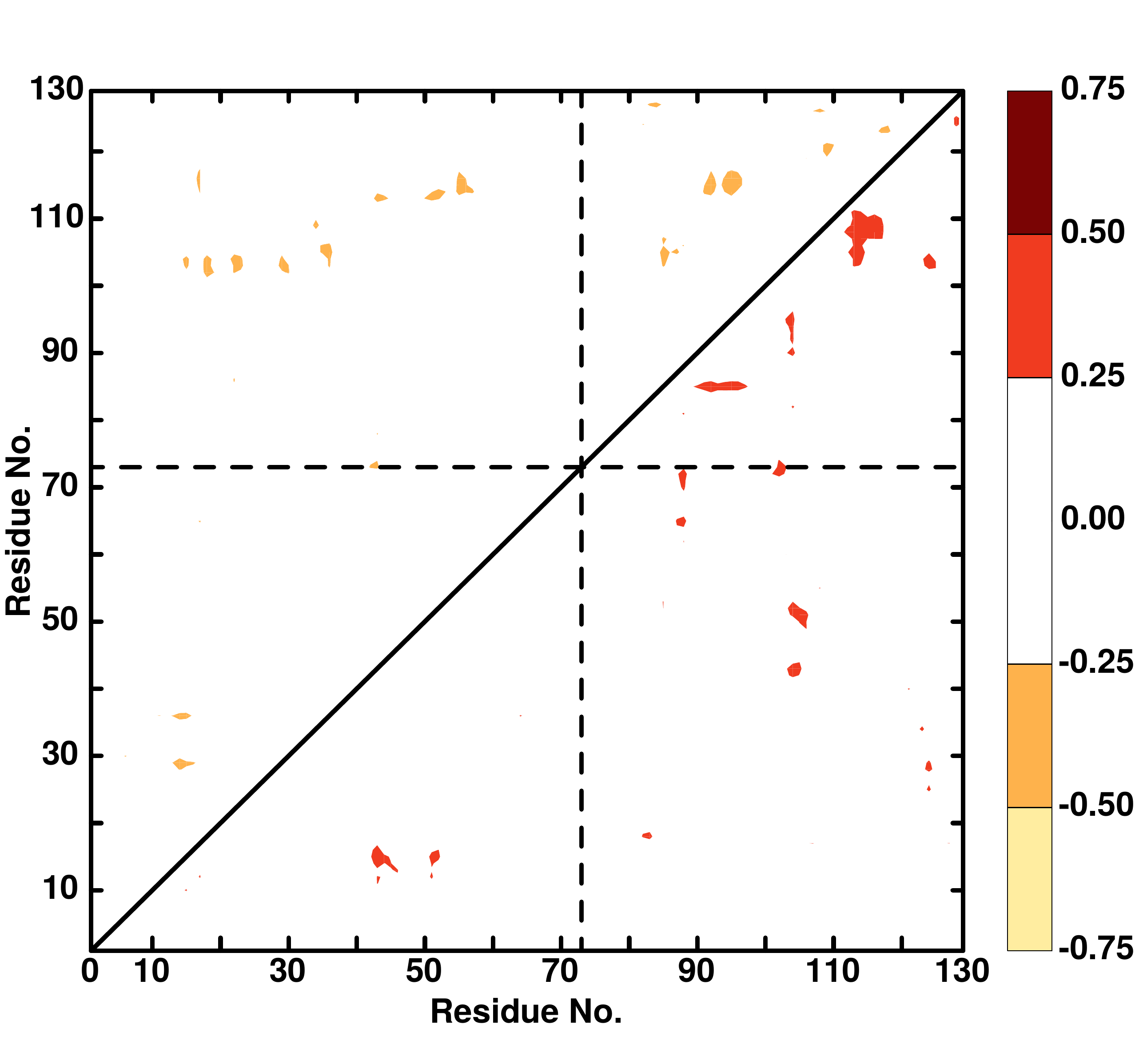}
\caption{$\Delta$DCCM between WT and Ala73N$_3$. Positive differences are in the lower right triangle, negative differences in the upper left triangle. Only differences with an absolute value greater than 0.25 are displayed.}
\label{sifig:diff73}
\end{center}
\end{figure}

\begin{figure}[H]
\begin{center}
\includegraphics[width=0.5\textwidth]{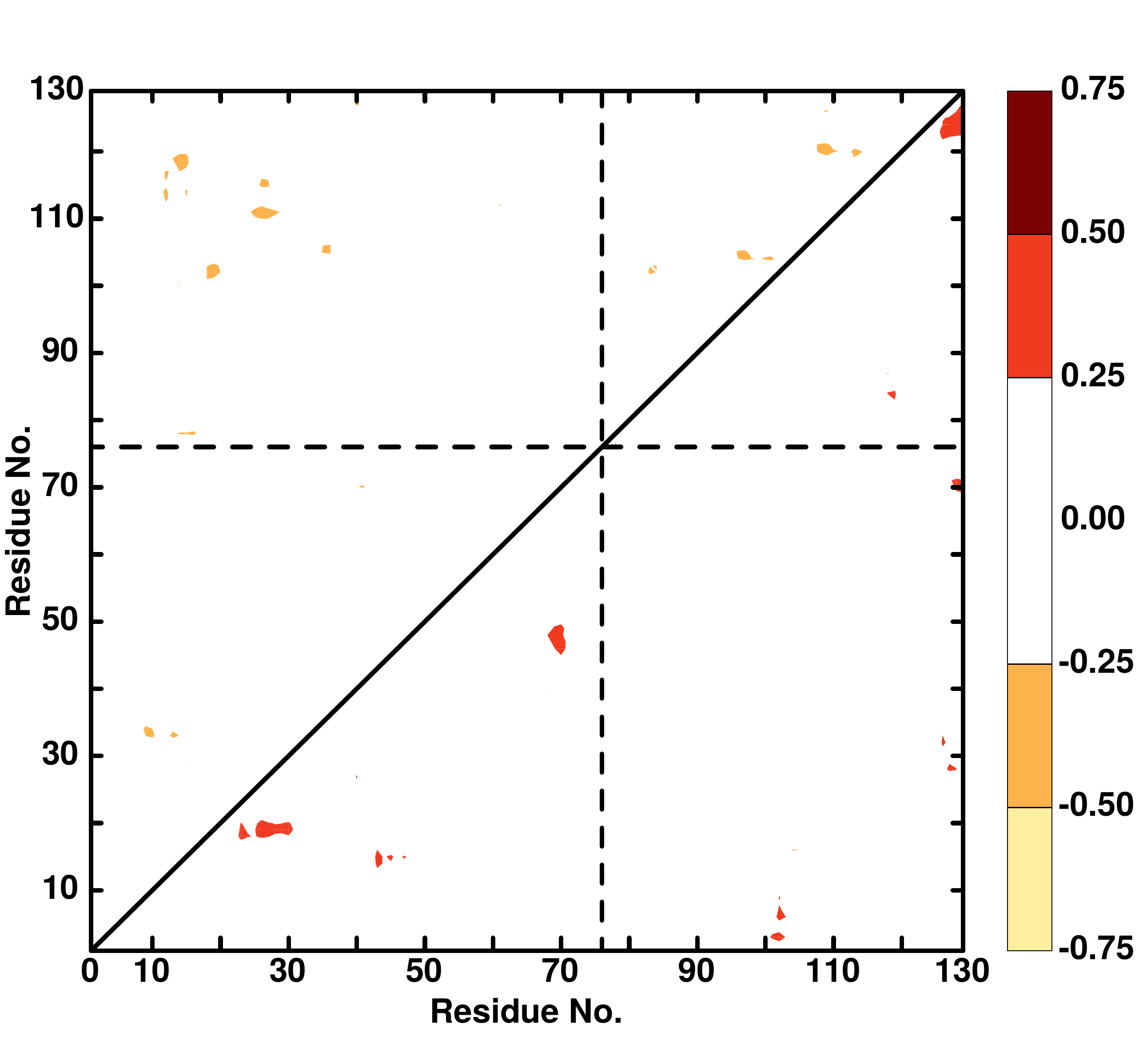}
\caption{$\Delta$DCCM between WT and Ala76N$_3$. Positive differences are in the lower right triangle, negative differences in the upper left triangle. Only differences with an absolute value greater than 0.25 are displayed.}
\label{sifig:diff76}
\end{center}
\end{figure}

\begin{figure}[H]
\begin{center}
\includegraphics[width=0.5\textwidth]{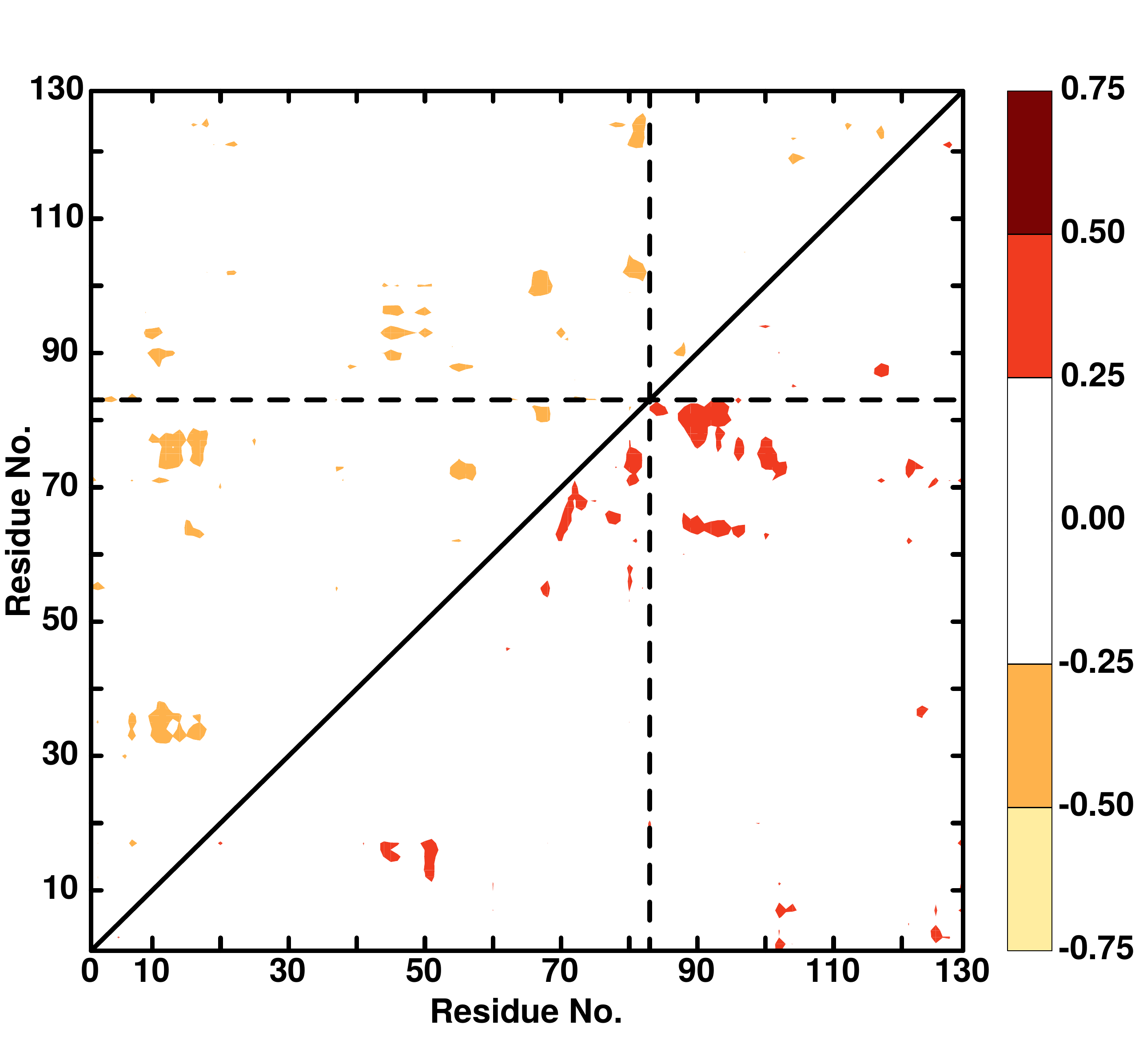}
\caption{$\Delta$DCCM between WT and Ala83N$_3$. Positive differences are in the lower right triangle, negative differences in the upper left triangle. Only differences with an absolute value greater than 0.25 are displayed.}
\label{sifig:diff83}
\end{center}
\end{figure}

\begin{figure}[H]
\begin{center}
\includegraphics[width=0.5\textwidth]{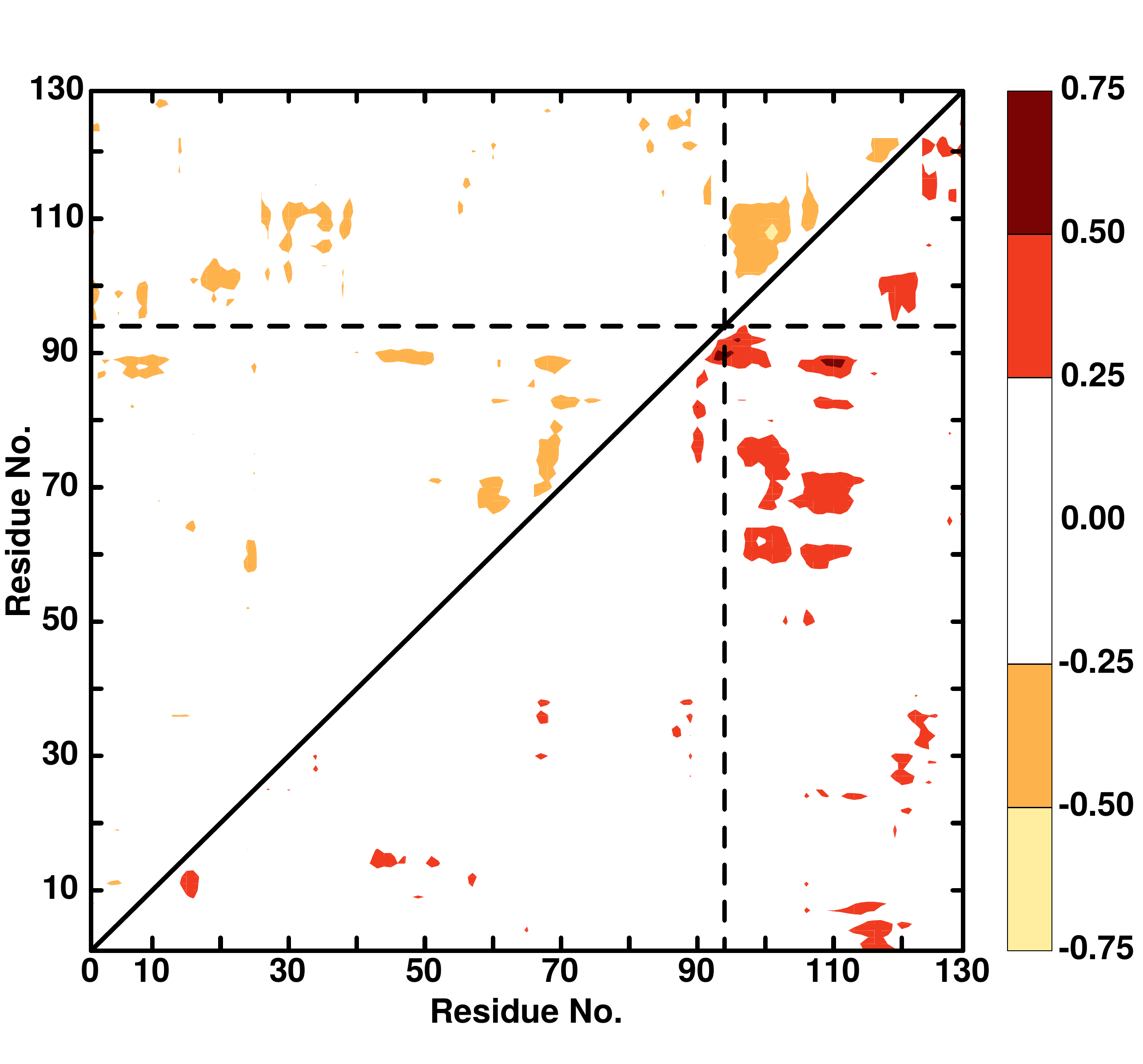}
\caption{$\Delta$DCCM between WT and Ala94N$_3$. Positive differences are in the lower right triangle, negative differences in the upper left triangle. Only differences with an absolute value greater than 0.25 are displayed.}
\label{sifig:diff94}
\end{center}
\end{figure}

\begin{figure}[H]
\begin{center}
\includegraphics[width=0.5\textwidth]{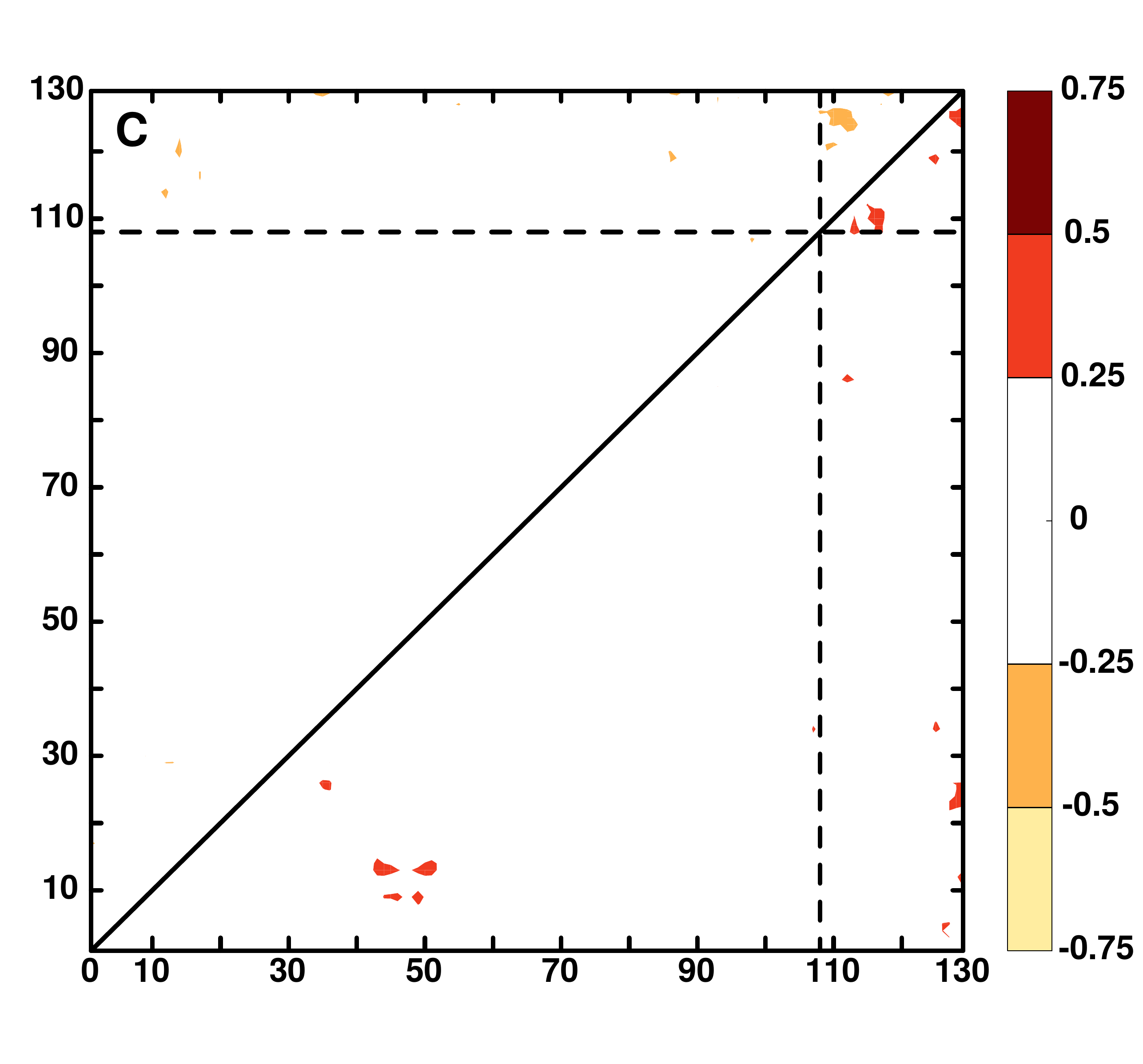}
\caption{$\Delta$DCCM between WT and Ala108N$_3$. Positive differences are in the lower right triangle, negative differences in the upper left triangle. Only differences with an absolute value greater than 0.25 are displayed.}
\label{sifig:diff108}
\end{center}
\end{figure}

\begin{figure}[H]
\begin{center}
\includegraphics[width=0.5\textwidth]{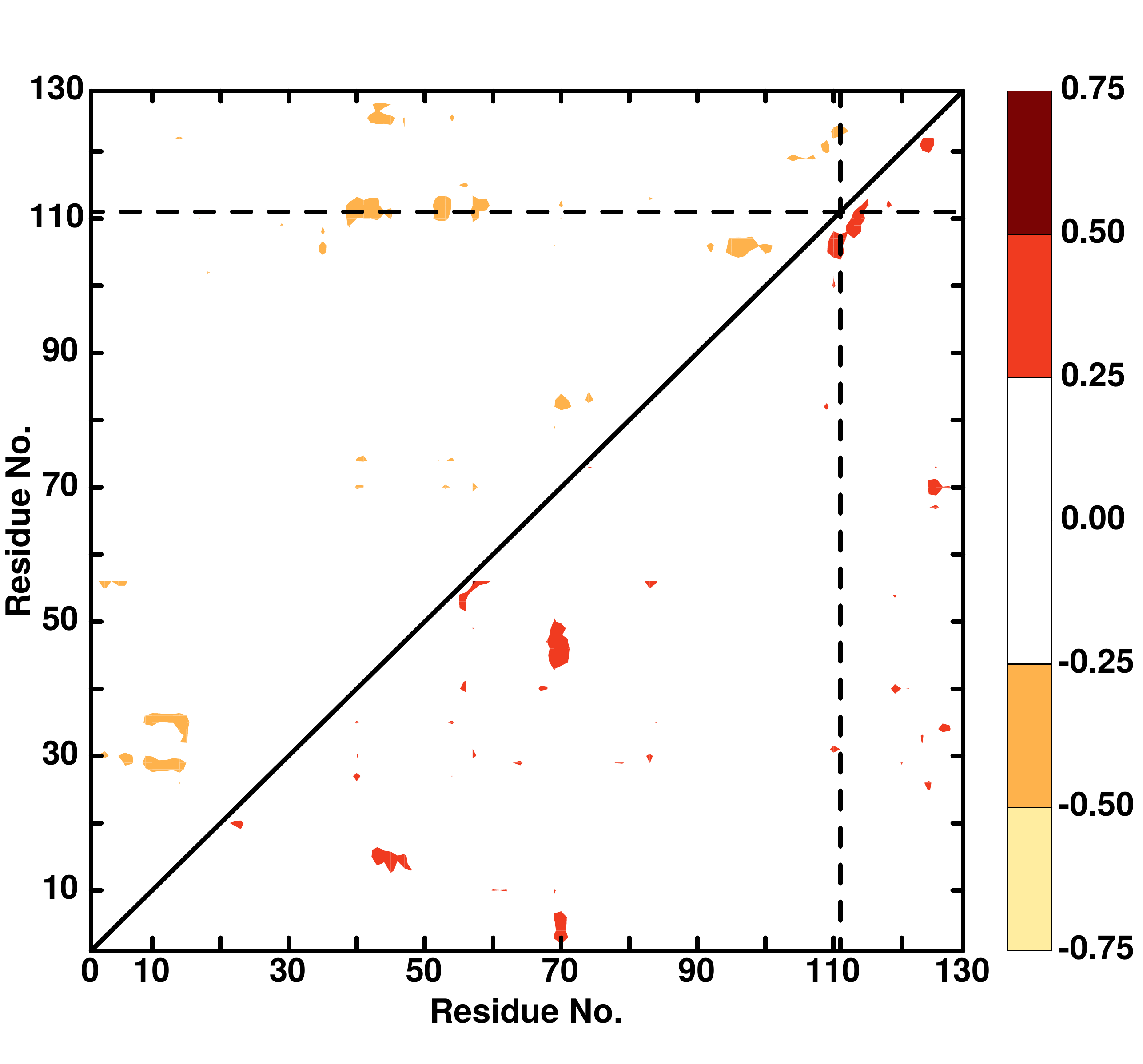}
\caption{$\Delta$DCCM between WT and Ala111N$_3$. Positive differences are in the lower right triangle, negative differences in the upper left triangle. Only differences with an absolute value greater than 0.25 are displayed.}
\label{sifig:diff111}
\end{center}
\end{figure}

\begin{figure}[H]
\begin{center}
\includegraphics[width=0.5\textwidth,angle=-90]{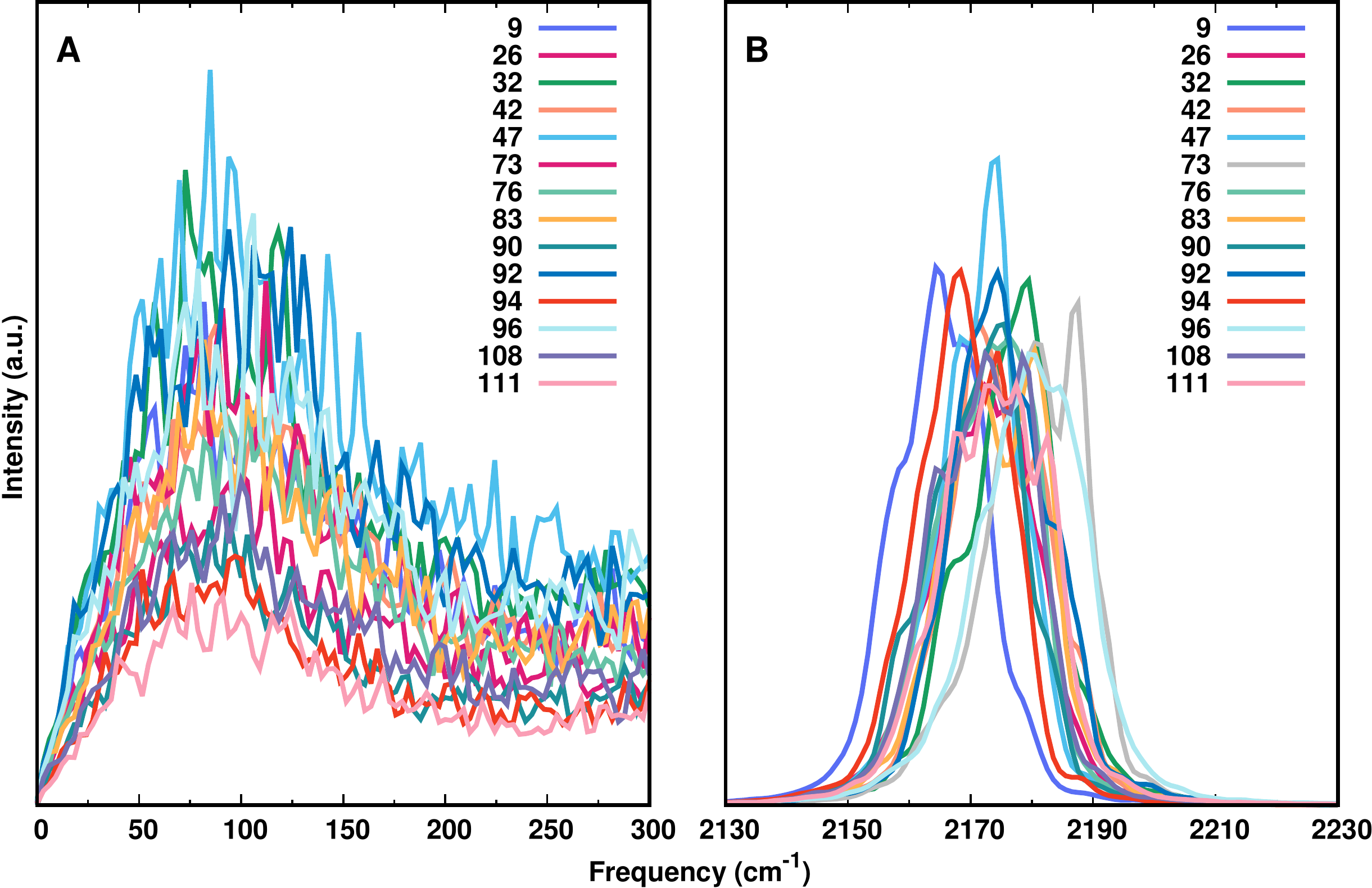}
\caption{IR spectrum obtained from Fourier Transform of the dipole moment auto-correlation function of protein (panel A) and -N$_3$ label (panel B) for all AlaN$_3$ modifications. 
The labels in each panel refer to the alanine residue
that carries the -N$_3$ moiety.}
\label{sifig:ir}
\end{center}
\end{figure}

\end{document}